\documentclass{PoS}
\usepackage{amsmath}
\usepackage{subfigure}

\newcommand{\gev}{\, {\rm GeV}}

\newcommand{\be}{\begin{equation}}
\newcommand{\ee}{\end{equation}}
\newcommand{\bea}{\begin{eqnarray}}
\newcommand{\eea}{\end{eqnarray}}

\title{Scalar and vector form factors of $D \to \pi \ell \nu$ and $D \to K \ell \nu$ decays with $N_f = 2+1+1$ Twisted fermions}

\ShortTitle{$D \to \pi \ell \nu$ and $D \to K \ell \nu$ form factors with $N_f=2+1+1$ Twisted fermions}

\author{ N. Carrasco$^{(a)}$, \speaker{P. Lami}$^{(a,b)}$, V. Lubicz$^{(a,b)}$, E. Picca$^{(a,b)}$,
L. Riggio$^{(a)}$, S. Simula$^{(a)}$, C. Tarantino$^{(a,b)}$
\\

\it $^{(a)}$ INFN, Sezione di Roma Tre, Rome, Italy. Email: \email{carrasco@fis.uniroma3.it}, \email{lorenzo.riggio@gmail.com}, \email{simula@roma3.infn.it}

\it $^{(b)}$ Dipartimento di Matematica e Fisica, Universit\`a  Roma Tre, Rome, Italy.  Email: \email{lamipaolo@gmail.com}, \email{lubicz@fis.uniroma3.it}, \email{e.picca88@gmail.com}, \email{tarantino@fis.uniroma3.it}
\\

\bf{For the ETM Collaboration}
}
       
\abstract{ We present a lattice QCD determination of the vector and scalar form factors of the semileptonic decays $D\to\pi\ell\nu$ and $D\to K\ell\nu$ which are relevant for the extraction of the CKM matrix elements $|V_{cd}|$ and $|V_{cs}|$ from experimental data. Our analysis is based on the gauge configurations produced by the European Twisted Mass Collaboration with $N_f = 2+1+1$ dynamical fermions. We simulated at three different values of the lattice spacing and with pion masses as small as $210$ MeV. Our preliminary estimates for the vector form factor at zero 4-momentum transfer are $f_+^{(D \to \pi)}(0) = 0.610 (23)$ and $f_+^{(D \to K)}(0) = 0.747 (22)$, where the uncertainties are only statistical. By combining our results with the experimental values of $f_+^{(D \to \pi)}(0) |V_{cd}|$ and $f_+^{(D \to K)}(0)|V_{cs}|$ we obtain $|V_{cd}| = 0.2336 (93)$ and $|V_{cs}| = 0.975 (30)$, which together with the PDG determination of $|V_{cb}|$ are in agreement with the unitarity constraint of the Standard Model.}

\FullConference{The 33rd International Symposium on Lattice Field Theory\\
		14 -18 July 2015\\
		Kobe International Conference Center, Kobe, Japan*}

\begin{document}

\section{Introduction and simulation details}

In the Standard Model (SM) weak charged flavor changing currents are regulated by the Cabibbo-Kobayashi-Maskawa (CKM) matrix \cite{Cabibbo:1963yz,Kobayashi:1973fv}.
Therefore, a precise knowledge of its elements allows us to test the SM and possibly to search for New Physics.

In this contribution we present the preliminary results of our analysis of the vector and scalar form factors of the $D \to \pi \ell \nu$ and $D \to K \ell \nu$ decays, which are relevant for the determination of the CKM matrix elements $|V_{cd}|$ and $|V_{cs}|$.

We use the ensembles of gauge configurations generated by the European Twisted Mass Collaboration (ETMC) with $N_f = 2+1+1$ dynamical quarks, which include in the sea, besides two light mass-degenerate quarks, also the strange and the charm quarks.
The gauge ensembles and the simulations are the same adopted in Ref.~\cite{Carrasco:2014cwa} to determine the up, down, strange and charm quark masses, as well as in Ref.~\cite{Carrasco:2014poa} to determine the leptonic decay constants $f_K / f_\pi$, $f_D$ and $f_{D_s}$.
We consider three different values of the lattice spacing, the smallest being approximately $0.06$ fm, in order to control properly the continuum extrapolation, and pion masses as low as $210$ MeV.

The gauge sector has been simulated using the Iwasaki gluonic action \cite{Iwasaki:1985we}, while sea quarks are implemented with the Wilson Twisted Mass Action \cite{Frezzotti:2003xj}, which at maximal twist allows for an automatic ${\cal{O}}(a)$-improvement \cite{Frezzotti:2003ni}. 
Valence quarks are simulated using the Osterwalder-Seiler action \cite{Osterwalder:1977pc}.
More details about the lattice ensembles and the simulation details can be found in Ref.~\cite{Carrasco:2014cwa}.
In order to inject momenta we implement non-periodic boundary conditions for the quark fields \cite{Bedaque,Tantalo,Guadagnoli:2005be}, obtaining in this way quark momenta ranging from $\simeq 150$ to $\simeq 650$ MeV.
This allowed us to cover the region of values of the squared 4-momentum transfer $q^2$ from $q^2 = 0$ to $q_{max}^2 = (M_D - M_{\pi(K)})^2$.  

We have calculated the three-point correlation functions for the two pseudoscalar mesons involved in the decays connected by the weak vector current in order to extract the corresponding vector current matrix elements.
From the latter we construct the form factors $f_+(q^2)$ and $f_0(q^2)$, which are then interpolated to the physical strange and charm quark masses, $m_s$ and $m_c$, determined in Ref.~\cite{Carrasco:2014cwa}, using a simple quadratic spline.
We perform a multi-combined fit of the lattice data by analysing simultaneously the dependencies of $f_+(q^2)$ and $f_0(q^2)$ on the (renormalized) light-quark mass $m_l$, the squared lattice spacing $a^2$ and the squared 4-momentum transfer $q^2$.

Our preliminary results at $q^2 = 0$ are $f_+^{(D \to \pi)}(0) = 0.610 (23)$ and $f_+^{(D \to K)}(0) = 0.747 (22)$, where the uncertainties are only statistical. 
By combining our results with the experimental values of $f_+^{(D \to \pi)}(0) |V_{cd}|$ and $f_+^{(D \to K)}(0) |V_{cs}|$ from Ref.~\cite{Amhis:2014hma} we obtain $|V_{cd}| = 0.2336 (93)$ and $|V_{cs}| = 0.975 (30)$, which together with the value $|V_{cb}| = 0.0413 (49)$ from the PDG \cite{PDG}, are in agreement with the unitarity constraint of the SM.

\section{Extraction of the vector and scalar form factors of the D-meson semileptonic decays}

In this section we present the extraction of the vector and scalar form factors describing the semileptonic decays of the D-meson from our lattice simulations.
We started by calculating the three-point correlation functions connecting the initial D-meson and the final pseudoscalar one (a pion or a kaon) through a (bare) local vector current $V_\mu$, employing Gaussian smearing for the meson interpolating fields at the source and the sink.
At large time distances one has
 \be
    \label{eq:3pointD}
    C_\mu^{D M}(t, T_{sep}, \vec{p}, \vec{p}^\prime) \xrightarrow{t \gg a ~ , ~ (T_{sep} - t) \gg a} \frac{\sqrt{Z_D(\vec{p}) Z_M(\vec{p}^\prime)}}{4E_D E_M} 
    \left< M(p^\prime) \left| V_{\mu} \right| D(p) \right> ~ e^{- E_D t - E_M (T_{sep} - t)} ~ ,
 \ee
where $T_{sep}$ is the time distance between the source and the sink, $M$ represents either a $\pi$ or a $K$ meson, $\left< M(p^\prime) \left| V_{\mu} \right| D(p) \right>$ is the (bare) vector current matrix element, $E_{D(M)}$ is the energy of the D(M)-meson and $Z_{D(M)}(\vec{p})$ is the coupling of the smeared interpolating field with the D(M)-meson.
The latter can be determined by analysing the large time distance behaviour of the two-point correlation functions, namely
 \be
    \label{eq:2pointD}
    C_2^{D(M)}(t, \vec{p} (\vec{p}^\prime)) \xrightarrow{t \gg a} \frac{Z_{D(M)}(\vec{p} (\vec{p}^\prime) )}{2E_{D(M)}} ~ \left[e^{- E_{D(M)} t } + e^{- E_{D(M)} (T - t)} \right].
 \ee

The matrix elements of the renormalized vector curent $\hat{V}_\mu$ are then given by \cite{Frezzotti:2003ni}
 \be
    \label{eq:hatV}
     \left< \hat{V}_{\mu} \right> \equiv \left< M(p^\prime) \left| \hat{V}_{\mu} \right| D(p) \right> = Z_V \left< M(p^\prime) \left| V_{\mu} \right| D(p) \right> ~ ,
 \ee
 where $Z_V$ is the renormalization constant of the vector current, determined very precisely by ETMC through the Ward-Takahashi identity in Ref.~\cite{Carrasco:2014cwa}. 
 From the matrix elements $\left< \hat{V}_{\mu} \right>$ we compute the vector and the scalar form factors according to the following relations:
 \bea
    \label{eq:fpf-}
    f_+(q^2) & = & \frac{(E_D - E_{K(\pi)}) \left< V_i \right> - (p_i - p_i^\prime) \left< V_0 \right>}{2E_D p_i^\prime - 2E_{K(\pi)} p_i} ~ , \\
    f_-(q^2) & = & \frac{(p_i + p_i^\prime) \left< V_0 \right> - (E_D + E_{K(\pi)}) \left< V_i \right>}{2E_D p_i^\prime - 2E_{K(\pi)} p_i} ~ , \\
    f_0(q^2) & = & f_+(q^2) + \frac{q^2}{M_D^2-M_{K(\pi)}^2} f_-(q^2)
 \eea

\begin{figure}[htb!]
\begin{center}
\includegraphics[scale=0.425]{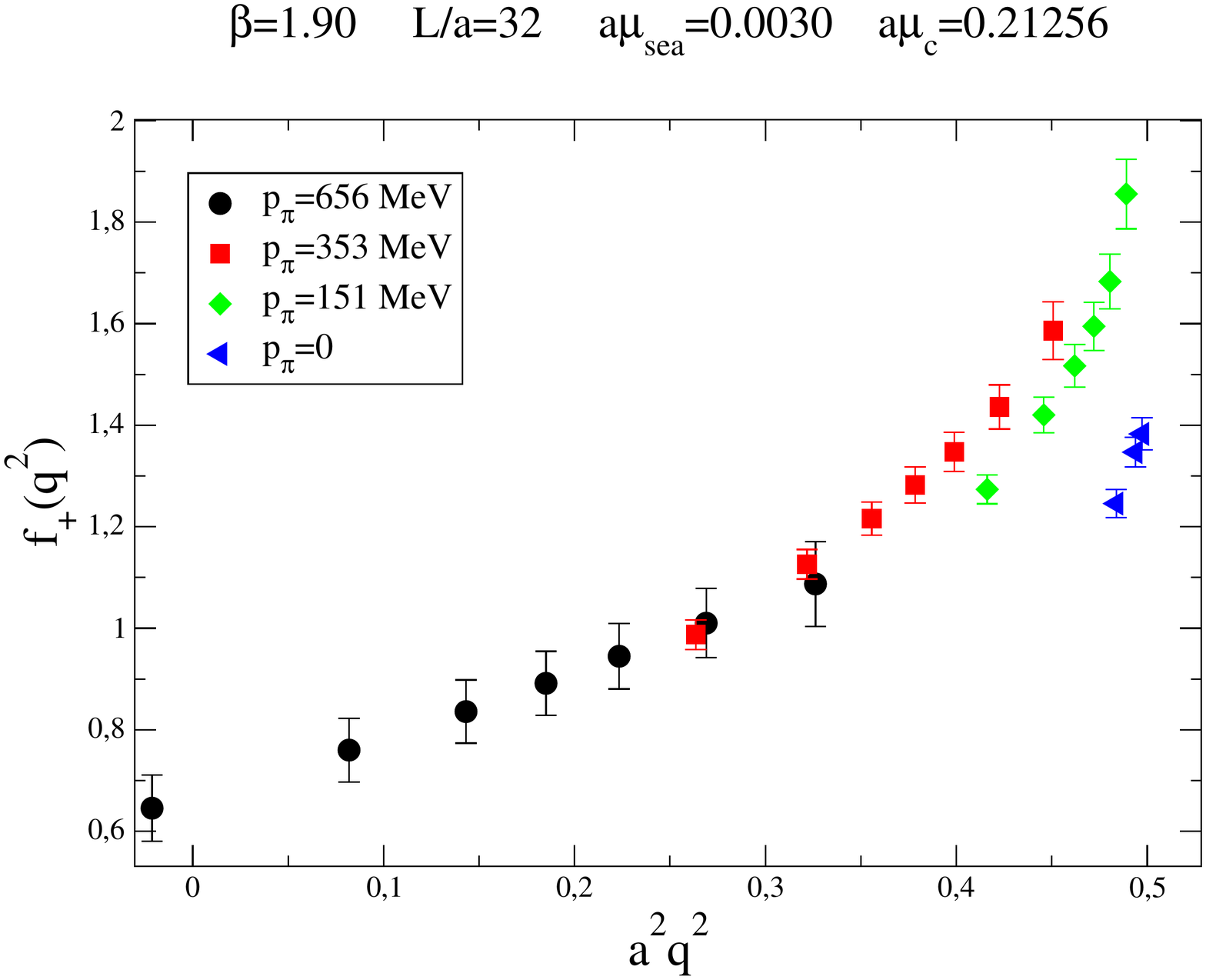}
\end{center}
\vspace*{-0.75cm}
\caption{Momentum dependence of the vector form factor $f_+(q^2)$ for the $D \to \pi \ell \nu$ decay. The data corresponds to the ensemble A30.32 (see Ref.~\protect\cite{Carrasco:2014cwa}) with $a \mu_\ell = 0.0030$ and $a \mu_c = 0.21256$. Different values of the pion momentum are shown by different markers and colors.}
\label{fig:fishbone}
\end{figure}

At fixed values of the quark masses and of the lattice spacing the $q^2$-dependence of the form factors exhibit the behavior shown in Fig.~\ref{fig:fishbone}.
Besides the dependence on the Lorentz-invariant quantity $q^2$ there is a clear dependence upon the value of the light-meson momentum. 
Such an effect turns out to be more relevant in the case of the $D \to \pi$ form factors with respect to the case of the $D \to K$ ones. 

\begin{figure}[htb!]
\begin{center}
\includegraphics[scale=0.425]{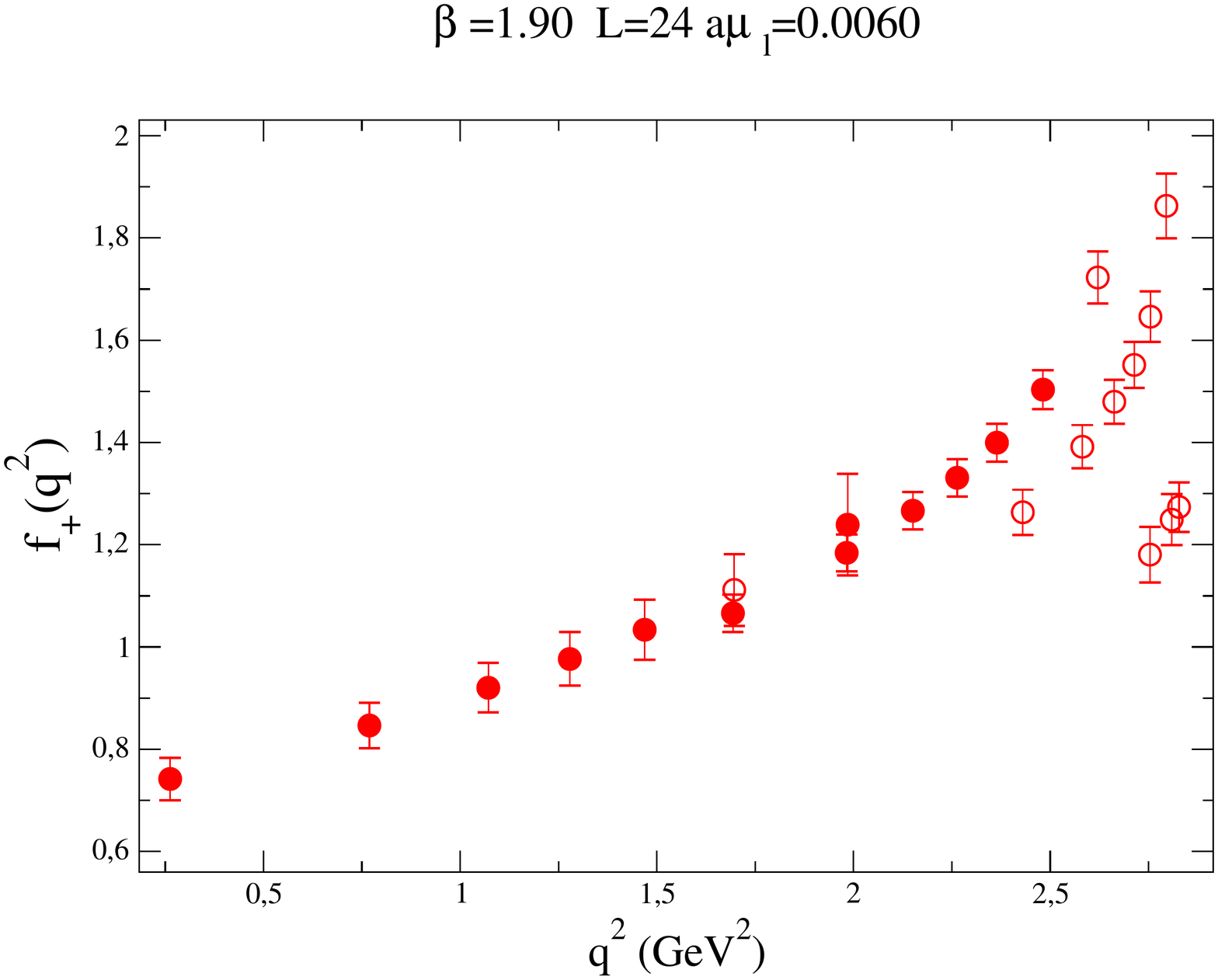}
\end{center}
\vspace*{-0.75cm}
\caption{Momentum dependence of the vector form factor $f_+(q^2)$ for the $D \to \pi \ell \nu$ decay corresponding to the ensemble A60.24 (see Ref.~\protect\cite{Carrasco:2014cwa}). The lattice data are interpolated to the physical charm quark mass determined in Ref.~\cite{Carrasco:2014cwa}. Hollow points represent the data excluded by the cut on the hypercubic invariant $\tilde{q}_E^4 / q_E^2$ (see text).}
\label{fig:fishbone_solved}
\end{figure}

The behavior shown in Fig.~\ref{fig:fishbone} is clearly due to the breaking of Lorentz invariance and the discretization effects responsible for that are expected to depend on hypercubic invariants and to be ${\cal{O}}(a)$-improved.
In terms of the quantities $q_E^2 \equiv \sum_{i=1}^4 q_i^2$ and $\tilde{q}_E^4 = \sum_{i=1}^4 q_i^4$ a possible hypercubic invariant of order ${\cal{O}}(a^2)$ is $a^4 \tilde{q}_E^4 / a^2 q_E^2$.
After looking at the dependence of the lattice data on $a^4 \tilde{q}_E^4 / a^2 q_E^2$ we have applied a cut by selecting only those data that satisfy the condition $\tilde{q}_E^4 / q_E^2 < 2.5 \gev^2$.
The effect of this cut is shown in Fig.~\ref{fig:fishbone_solved}, where the full red dots represent the data passing the cut condition.
It is clear that after the cut the lattice data appear to depend smoothly on the Lorentz-invariant quantity $q^2$.

\section{Combined chiral and continuum extrapolations}

We now perform a global fit of the lattice data of the vector and scalar form factors by analysing simultaneously their dependencies on $m_\ell$, $a^2$ and $q^2$.
We adopt a simple polar expression with polynomial corrections in $m_\ell$, $a^2$ and $q^2$, viz.
 \bea
    \label{eq:dpikfit}
    f_+(q^2) & = & \frac{F_+}{1 - q^2 / M_V^2} (1 + A q^2) (1 + P_1m_\ell + P_2 a^2) ~ , \\
    f_0(q^2) & = & \frac{F_+}{1 - q^2 / M_S^2} (1 + B q^2) (1 + P_1m_\ell + P_2 a^2) ~ ,
 \eea
where $F_+$, $A$, $B$, $P_1$ and $P_2$ are free parameters and $M_V$($M_S$) is the mass of the low-lying vector (scalar) resonance, i.e.~the $D^*(2010)$ and $D_0^*(2400)$ resonances, respectively. The values of $M_V$ and $M_S$ are estimated as $M_{V(S)} = M_D + \Delta_{V(S)}$, where $M_D$ is the $D$-meson mass calculated on the lattice for each gauge ensemble, while $\Delta_{V(S)}$ is the difference between the experimental value of the low-lying vector (scalar) resonance mass and the isospin symmetric D-meson mass, taken from the PDG \cite{PDG}. 

\begin{figure}
\begin{center}
\includegraphics[scale=0.55]{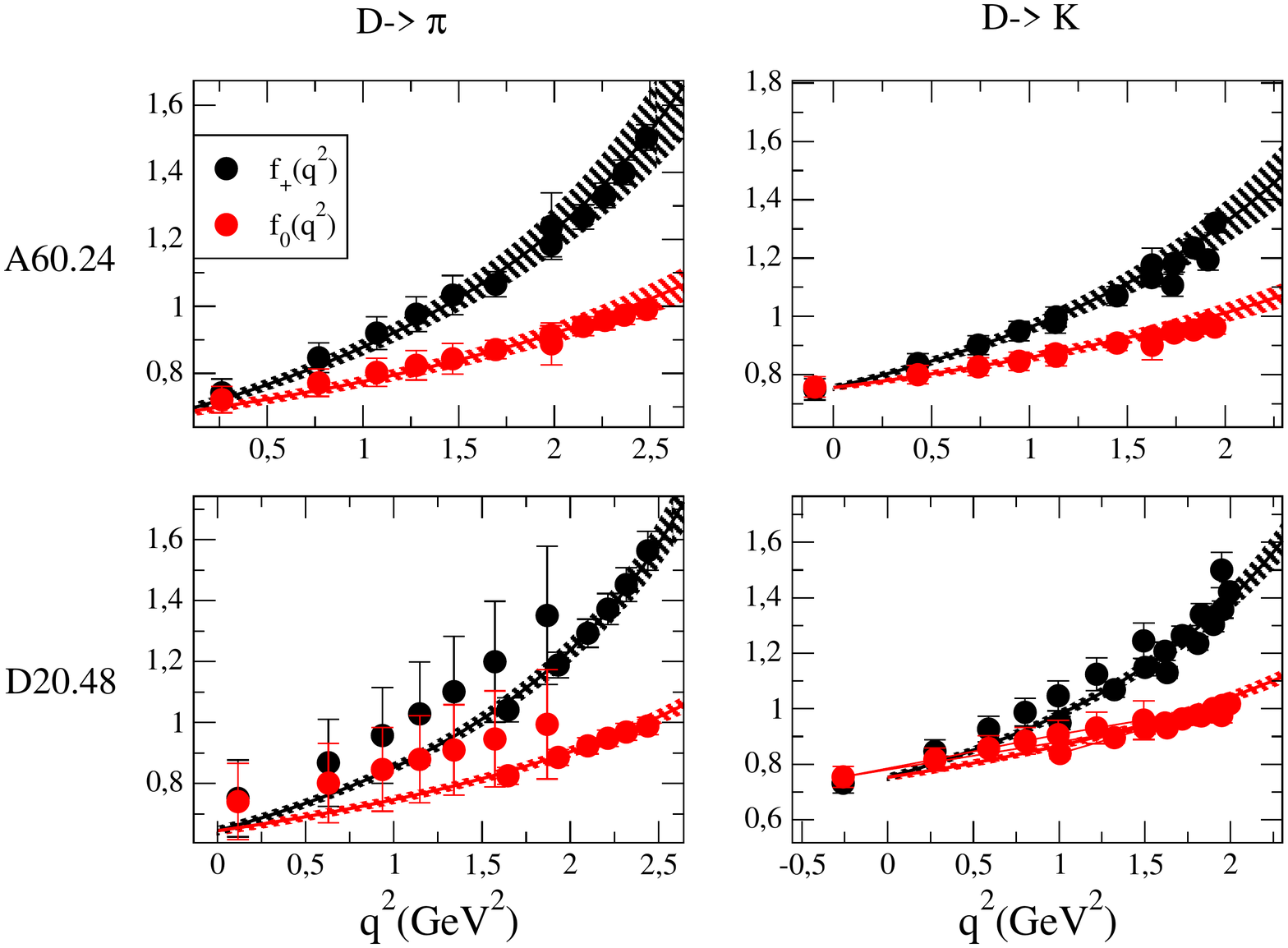}
\end{center}
\vspace*{-0.5cm}
\caption{Data points of $f_+(q^2)$ and $f_0(q^2)$ for the $D \to \pi \ell \nu$ decay (left panel) and for the $D \to K\ell \nu$ decay (right panel) corresponding to the gauge ensembles A60.24 and D20.48 (see Ref.~\protect\cite{Carrasco:2014cwa}). The solid curves represent the results of the fit (\protect\ref{eq:dpikfit}) together with the uncertainties indicated by the slashes.}
\label{fig:dpidk_ens}
\end{figure}

The comparison between the lattice data and the results of the fitting functions (\ref{eq:dpikfit}) obtained in the case of both $D \to \pi \ell \nu$ and $D \to K \ell \nu$ is reported in Fig.~\ref{fig:dpidk_ens} in the case of the ETMC ensembles A60.24 and D20.48 (see Ref.~\protect\cite{Carrasco:2014cwa}).

\begin{figure}
\begin{subfigure}{}
\hspace{-0.5cm}
\includegraphics[scale=0.30]{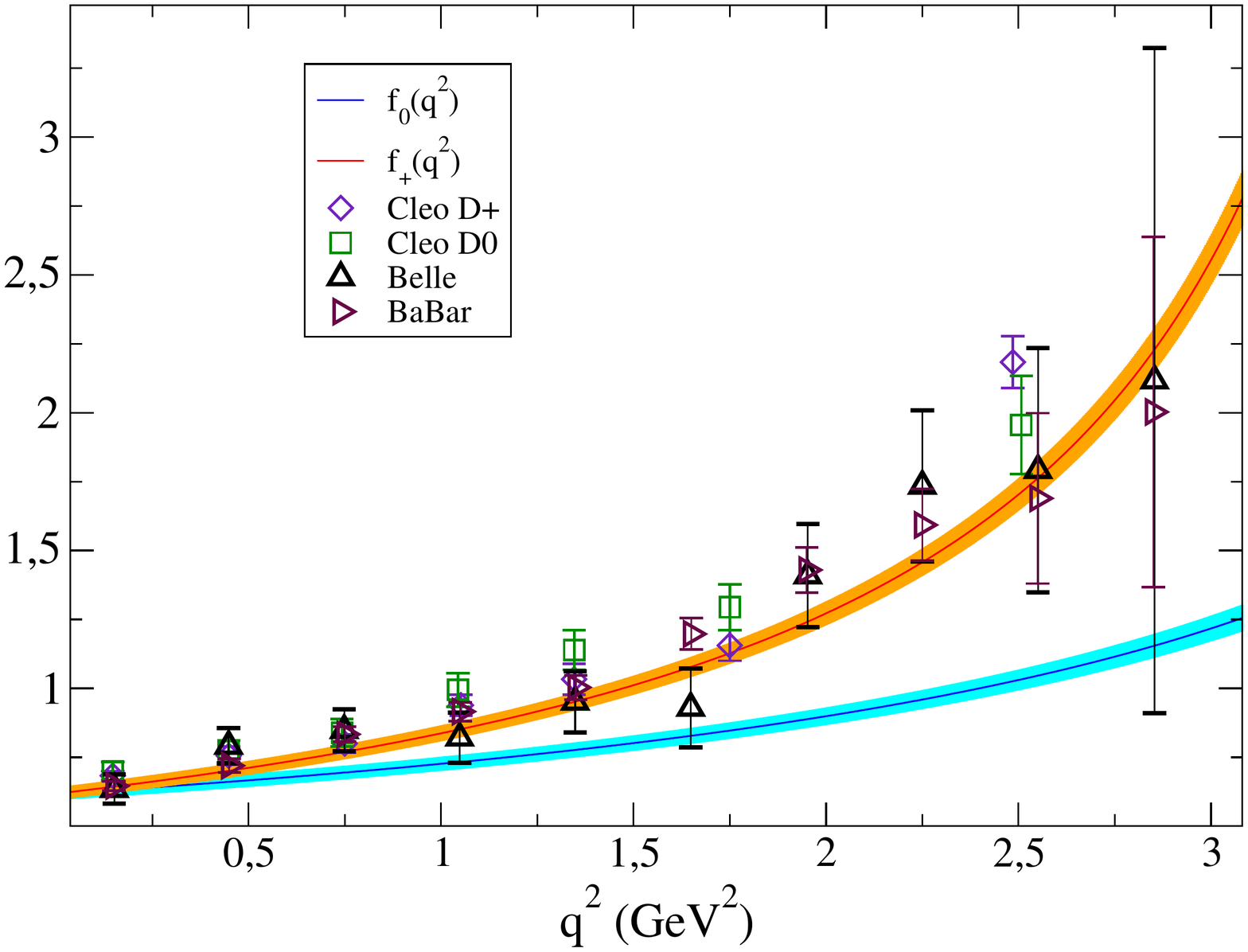}
\end{subfigure}
\begin{subfigure}{}
\hspace{-1.5cm}
\includegraphics[scale=0.30]{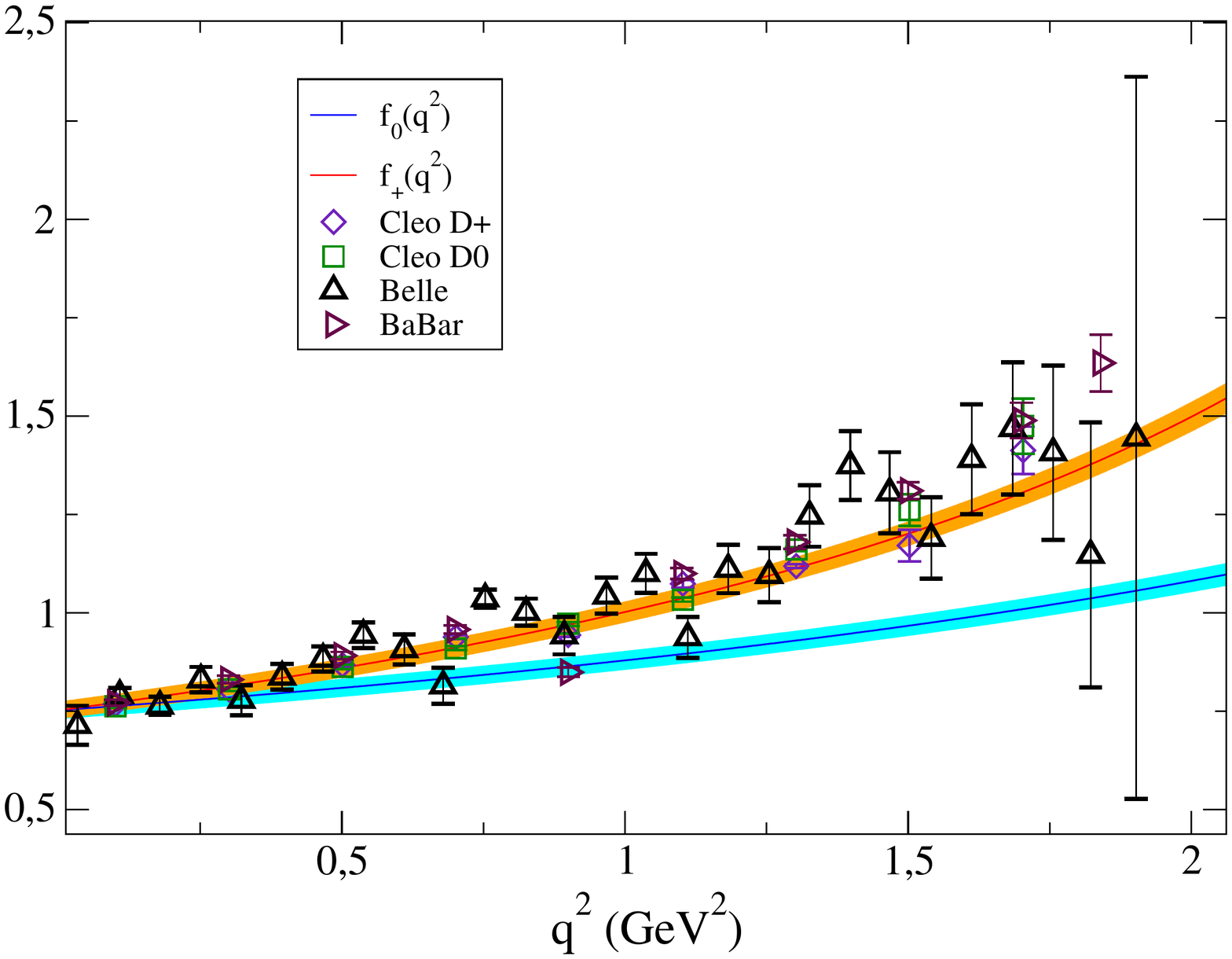}
\end{subfigure}
\vspace*{-0.5cm}
\caption{Results for the vector and scalar form factors of the $D \to \pi \ell \nu$ (left panel) and $D \to K \ell \nu$ decays extrapolated to the physical point as functions of $q^2$. The results of the Belle \cite{Widhalm:2006wz}, Babar \cite{Lees:2014ihu,Aubert:2007wg} and Cleo \cite{Besson:2009uv,Dobbs:2007aa} experiments for the vector form factor $f_+(q^2)$  are reported for comparison.}
\label{fig:dpikres}
\end{figure}

In Fig.~\ref{fig:dpikres} our results for the form factors extrapolated to the physical point are compared with the experimental data for the vector form factor $f_+(q^2)$ obtained by the Belle \cite{Widhalm:2006wz}, Babar \cite{Aubert:2007wg,Lees:2014ihu} and Cleo \cite{Besson:2009uv,Dobbs:2007aa} experiments.
It can clearly be seen that the agreement is good.

Our preliminary estimates for the vector form factor at zero 4-momentum transfer are
 \bea
    \label{eq:dpikres}
    f_+^{(D \to \pi)}(0) & = & 0.610 ~ (23) ~ , \\[2mm]
    f_+^{(D \to K)}(0) & = & 0.747 ~ (22) ~ ,
 \eea
where the uncertainties are only statistical. 
Our results can be compared with the FLAG averages $f_+^{(D \to \pi)}(0) = 0.666 (29)$ and $f_+^{(D \to K)}(0) = 0.747 (19)$ \cite{FLAG2}, based on the lattice results obtained at $N_f = 2+1$ in Refs.~\cite{Na:2011mc} and \cite{Na:2010uf}, respectively.

\section{Calculation of $|V_{cd}|$ and $|V_{cs}|$ }

The results (\ref{eq:dpikres}) for the form factor at zero 4-momentum transfer can be combined with the updated experimental averages $f_+^{(D \to \pi)} (0)|V_{cd}| = 0.1425 (19)$ and $f_+^{(D \to K)}(0) |V_{cs}| = 0.728 (5)$ from Ref.~\cite{Amhis:2014hma} to get
 \bea
    |V_{cd}| & = & 0.2336 ~ (31)_{exp} ~ (88)_{f_+(0)} = 0.2336 ~ (93) ~ , \\[2mm]  
    |V_{cs}| & = & 0.975 ~ (7)_{exp} ~ (29)_{f_+(0)} = 0.975 ~ (30) ~ .
 \eea
These results, together with the latest determination of $|V_{cb}| = 0.0413 (49)$ from the PDG \cite{PDG}, can be used to perform the unitarity test of the second row of the CKM matrix, obtaining
 \be
    \label{eq:ckmunit2ndrow}
    |V_{cd}|^2 + |V_{cs}|^2 + |V_{cb}|^2 = 1.007 ~ (63) ~ ,
 \ee
which agrees with the SM constraint at the level of $\simeq 6\%$.

\section*{Acknowledgements}

\noindent We acknowledge the CPU time provided by the PRACE Research Infrastructure under the project PRA067 ``First Lattice QCD study of B-physics with four flavors of dynamical quarks'' awarded at JSC (Germany), and by the agreement between INFN (Italy) and CINECA (Italy) under the specific initiative INFN-LQCD123.
V.~L., S.~S.~and C.~T.~thank MIUR (Italy) for partial support under Contract No. PRIN 2010-2011.
L.~R.~thanks INFN (Italy) for the support under the SUMA computing project (https://web2.infn.it/SUMA).

\end{document}